# The desalting/salting pathway: a route to form metastable aggregates with tuneable morphologies and lifetimes

Sumit Mehan[a], Laure Herrmann[b], Jean-Paul Chapel[c], Jacques Jestin[a], Jean-Francois Berret[b], Fabrice Cousin*[a]

We investigate the formation/re-dissociation mechanisms of hybrid complexes made from negatively charged $PAA_{2k}$ coated $\gamma\text{-}Fe_2O_3$ nanoparticles (NP) and positively charged polycations (PDADMAC) in aqueous solution in the regime of very high ionic strength (I). When the building blocks are mixed at large ionic strength (1M $NH_4Cl$), the electrostatic interaction is screened and complexation does not occur. If the ionic strength is then lowered down to a targeted ionic strength $I_{target}$, there is a critical threshold $I_c$ = 0.62 M at which complexation occurs, that is independent on the charge ratio Z and the pathway used to reduce salinity (drop-by-drop mixing or fast mixing). If salt is added back up to 1M, the transition is not reversible and persistent out-of-equilibrium aggregates are formed. The lifetimes of such aggregates depends on $I_{target}$: the closer $I_{target}$ to $I_c$ is, the more difficult it is to dissolve the aggregates. Such peculiar behavior is driven by the inner structure of the complexes that are formed after desalting. When $I_{target}$ is far below $I_c$, strong electrostatic interactions induce the formation of dense, compact and frozen aggregates. Such aggregates can only poorly reorganize further on with time, which makes their dissolution upon resalting almost reversible. Conversely, when $I_{target}$ is close to $I_c$ more open aggregates are formed due to weaker electrostatic interactions upon desalting. System can thus rearrange with time to lower its free energy and reach more stable out-of-equilibrium states which are very difficult to dissociate back upon resalting, even at very high ionic strength.

## Introduction

Non-equilibrium assemblies show various interesting and time-dependent properties that are not accessible to systems at equilibrium, the most prominent examples being living systems. [1,2,3] The possibility to realize controlled time dependent features provide these systems both complexity and potential for applications in designing new materials with dedicated properties.[4,5,6,7] These systems are either dissipative assemblies that require constant supply of energy[8] or pathway dependent kinetically trapped assemblies. The latter are molecular assemblies that have reached free energy local minima when exploring phase space along a given pathway, but are still away from the true minimum (equilibrium structure). [4,9,2] Depending on the energy landscape, a large number of local minima's are potentially accessible, and consequently different macromolecular assemblies with various morphologies and properties may be obtained from the same chemistry and predefined building blocks with the same final solution conditions.[10,11] If a specific property of the system that corresponds to a local minima is targeted, the pathway has to be chosen properly to reach such minima. [4,12,13,14] Usually the molecular assemblies are kinetically trapped when strong non-covalent interactions drive the assembly process into the kinetically trapped metastable states far from the true minimum of the free energy.[6,4,15,16,17,18,19]

Complexes of oppositely charged polyelectrolytes (PECs) are one of the interesting examples of kinetically controlled assemblies in solution[20,21,22] but also at interfaces.[23] The complexes of oppositely charged synthetic polyelectrolytes, which are excellent model to depict biopolymers systems, show diverse phase behaviour depending on the charge ratio of oppositely charged polyelectrolytes, molecular weight and mixing pathway. [24,16,25,26,27,28,29,30] In such system, there exists a critical threshold ionic strength $I_c$ above which electrostatic attractions between species of opposite charges are so screened that complexation no longer operates. [24,31,32, 33,34]

The presence of such threshold enables to drive the complexation of these structures using a desalting pathway. Starting from a mixture of oppositely charged polyelectroytes in a highly salted ''dormant'' solution where the building-blocks are not interacting anymore, it is then possible to trigger complexation by lowering the ionic strength below $I_c$, *i.e.* the electrostatic interaction can be switched *on* and *off* by tuning the ionic strength.

Such desalting route strategy is not limited to PECs and can be used with complexes of inorganic charged nanoparticles (NPs) and polyelectroytes of opposite charges to design different functional materials.[34,35,36,37,18] $I_c$ depends on several parameters (polyelectroyte length[38], particle size[39,40], polyelectroyte stiffness[41], or volume ratio of charged species due to ion-pairing[42]) and is usually larger than 0.1M. This route has for example already been used to form stable spherical large complexes of magnetic nanoparticles of maghemite $\gamma$-$Fe_2O_3$ coated with negatively charged $PAA_{2k}$ polyelectrolyte (polyacrylic acid) and positively charged PDADMAC (poly(diallyldimethylammonium chloride)) polyelectrolytes by crossing $I_c$ by dialysis.[35,36] Such complexes are further resistant to dilution. When the dialysis is performed in the presence of a magnetic field to cross the desalting transition, nano-rods are obtained instead of spheres due to magnetic dipolar interactions between magnetic NPs in the system. The thickness and length of these magnetic nano-rods can be modified by tuning the formulation conditions. These magnetic nano-rods are superparamagnetic, biocompatible and can be functionalized with nano-dots for fluorescent properties, providing them a huge potential in biomedical applications. Remarkably, they keep their anisotropic shape out of magnetic field and do not reorganize themselves into spherical objects, suggesting that they are frozen structures. The desalting pathway can be used as well between oppositely charged NPs ($CeO_2$) in solution and at interfaces.[43]

In the present work, we study the impact of desalting/resalting pathway on mixtures of oppositely charged nanoparticles and polymers. Shortly, the idea is to monitor the evolution of the structure of the system along the formulation pathway : *(a)* start from a "dormant" solution made from a building-block mixture at a very high ionic strength far above $I_c$; *(b)* cross the transition by desalting down to a targeted ionic strength $I_{target} < I_c$; *(c)* let the system evolve at $I_{target}$ for a given time, (d) and cross back the transition by resalting back up to the nominal ionic strength. Owing to the potential for applications of the PDADMAC/ PAA@$\gamma$-$Fe_2O_3$ system, we have chosen to investigate the very same experimental system. Moreover, as it has been shown in literature that different mixing pathways (*e.g.* fast *versus* drop by drop mixing) may lead to different quenched morphologies, we will assess if the mixing pathway does influence the final morphology of the quenched complexes. We address the following questions: *(i)* Does $I_c$ depends on Z (charge ratio between oppositely charged species) and/or on the formulation pathway ? *(ii)* Is the desalting transition perfectly reversible ? *(iii)* Can this system be utilized to form frozen assemblies ? *(iv)* What is the internal structure of frozen assemblies?

## Materials and Methods

### Materials

The sodium salt of poly(acrylic acid) ($M_W$ = 2100 g/mol) with Batch No. 420344 and poly(diallyldimethylammonium chloride) (PDADMAC) ($M_W$ < 100,000 g/mol) with Batch No. 522376 were purchased from Aldrich. The iron oxide nanoparticles ($\gamma$-$Fe_2O_3$) were synthesized by Massart method[44] by alkaline co-precipitation of iron(II) and iron(III) salts. At the end of synthesis, the nanoparticles were positively charged which ensures colloidal stability at low pH (~2) (>1 year) due to a combination of electrostatic and short range hydration repulsions and their counterions were $HNO_3^-$. The nanoparticles were then coated with small poly(acrylic acid) (PAA) chains using precipitation-redispersion protocol.[45,46] Since the $pK_a$ of PAA chains is around 4.7, coated $\gamma$-$Fe_2O_3$ nanoparticles, hereafter named PAA@$\gamma$-$Fe_2O_3$, are negatively charged at neutral and high pH, and are thus stabilized by electro-steric interactions and have stability for years around neutral pH. The characterization, coating properties and stabilization of these particles are described in previous publications.[46] In particular, the charge on nanoparticles have been measured by titration measurements. The size distribution of the bare $\gamma$-$Fe_2O_3$ nanoparticles follow a lognormal distribution with a mean diameter $R_{mean}$ = 4.7nm ($R_0$ = 4.5 nm and $\sigma$ = 0.2), as determined by SAXS. The nanoparticles are organized in solution as small branched clusters of ~ 4 NPs, modelled by the form factor of a pearl neckless (see Figure S1 in supplementary information).[47]

### Sample preparation

The charge ratio Z of PDADMAC to nanoparticles is calculated as described in reference[46] from the charges of nanoparticles (titration measurements) and the $M_W$ and charges of the PDADMAC chains.

### Preparation of initial stock solution and mixing protocol.

A stock solution of 0.05 v/v% $\gamma$-$Fe_2O_3$ at 1 M $NH_4Cl$ and pH 8 is prepared from a stock solution of $\gamma$-$Fe_2O_3$ nanoparticles at 1.15 wt% $\gamma$-$Fe_2O_3$ by dilution using deionized milli-Q water. A stock solution of PDADMAC at 1 M $NH_4Cl$ and pH 8 was prepared separately and its concentration was varied in order to reach the targeted Z after mixing. The ionic strength and pH of solutions were set to 1 M $NH_4Cl$ and pH 8 using 4M $NH_4Cl$, $NH_4OH$ and HCl solutions. The $NH_4Cl$ salt was chosen to allow comparison with complexation processes of very close systems already probed in literature (pure PAA2K/PDADMAC or PAA@CeO2 NPs/PDADMAC)[23,46]. pH 8 was chosen in order to ensure that PAA chains are fully charged. The equal volumes of 2Z-PDADMAC + 1 M $NH_4Cl$ and 0.05 v/v% $\gamma$-$Fe_2O_3$ + 1 M $NH_4Cl$ solutions were then mixed together to form the non-reactive clear solution, i.e. the so-called "dormant" solution of 0.025 v/v% $\gamma$-$Fe_2O_3$/Z-PDADMAC + 1 M $NH_4Cl$.

### Desalting transition pathway protocols

Three different charge ratios Z of respectively 0.004, 0.2 and 1 were investigated in the study. Two pathways were assessed to lower the ionic strength of the dormant solution at a given Z, hereafter named respectively the quenching and the drop-by-drop mixing pathway.

For the *quenching* pathway, the stock dormant solution (0.025 v/v% $\gamma$-Fe$_2$O$_3$/Z-PDADMAC at 1 M NH$_4$Cl) was diluted 10 times in one shot using a syringe or a pipette to correspond to the final targeted ionic strength I$_{target}$ in the range (0.1-1 M NH$_4$Cl). Samples were shaken gently by hand to ensure homogenization.

For the *drop-by-drop* mixing pathway, the stock dormant solution (0.025 v/v% $\gamma$-Fe$_2$O$_3$/Z-PDADMAC at 1 M NH$_4$Cl) was diluted 10 times drop by drop using a syringe filled with milli-Q water. After each drop, the solution was stirred to achieve a proper mixing. Typically, 100 drops were added sequentially to reach the targeted ionic strength. Drops were added every 3 seconds to achieve a characteristic time around 5 minutes for the whole process.

**Resalting pathway protocols**

The resalting pathway consists in diluting a desalted sample with a very concentrated solution of NH$_4$Cl to increase back the overall ionic strength to its initial value of 1 M NH$_4$Cl. It has been performed on samples that have been desalted by either the quenching or the drop-by-drop pathway after a specified time. For DLS measurements, the resalting was performed 1h or 30h after desalting. For the SAXS measurements, the resalting was performed 10 days after desalting.

In practice, 75 v/v% of the desalting solution (0.0025 v/v% $\gamma$-Fe$_2$O$_3$/Z-PDADMAC at *x* M NH$_4$Cl, where *x* < 1) is diluted with 25v/v% of corresponding solution at high ionic strength (*y* M NH$_4$Cl, where *y* ≤ 4) to set the NH$_4$Cl concentration back to 1 M NH$_4$Cl. This thus reduces the overall concentration of nanoparticles. In order to recover the nominal concentration of nanoparticle for the DLS measurements, we took benefit from the fact that the nanoparticles aggregate and settle at the bottom with time after desalting (see results part). That results in a two-phase system with solid precipitates and a supernatant solution almost free of nanoparticles. We removed 25v/v% of the supernatant from the desalted solution, and added a similar volume of high concentration *y* M NH$_4$Cl solution to increase the ionic strength to 1 M NH$_4$Cl.

For the SAXS measurements, a rather high concentration of nanoparticles was mandatory to obtain a good signal-to-noise ratio, which prompted us to modify the protocol. 5 ml of de-salting solution of 0.0025 v/v% $\gamma$-Fe$_2$O$_3$/Z-PDADMAC at *x* M NH$_4$Cl was prepared and was kept during 10 days for ageing. This time was sufficient for the nanoparticles to sediment. 4.5 ml of supernatant was then removed from this desalting solution to increase its concentration of nanoparticles by approximately 10 times. The 0.5 ml resultant solution was then gently shaken to make it homogeneous. Then, 0.125 ml of the solution was removed and an equivalent volume at *y* M NH$_4$Cl was added to set the final concentration to 1 M NH$_4$Cl.

**Small-angle X-ray scattering**

Small-angle X-ray scattering experiments were carried out with an Xeuss 2.0 instrument from Xenocs (Grenoble, France). It uses a micro-focused sealed tube Cu K-alpha source with a wavelength of 1.54 Å. The experiments were performed with collimated beam size of 0.5 x 0.6 mm. The sample to detector distance was set to 2.5 m to achieve a Q range from 0.045 nm$^{-1}$ to 0.25 nm$^{-1}$. The solutions were poured (0.025 v/v% $\gamma$-Fe$_2$O$_3$/Z-PDADMAC) inside 1.5 mm glass capillaries. The measurements were performed for 30 minutes to achieve a good statistics on the whole Q range. The respective scattering from capillaries, solvent and dark count were subtracted using standard protocols. The data were normalized to absolute units in cm$^{-1}$.

For the samples showing a macroscopic phase separation (sedimentation), we checked that the structure of the complexes was similar all along the capillary, except for the concentration factor. In order to optimize statistics, scattering curves were then taken at the bottom of the capillary where the complexes concentration was the highest.

The respective electronic scattering densities of the components are respectively $\rho_e(\gamma$-Fe$_2$O$_3) = 39.42 \; 10^{10}$ cm$^{-2}$, $\rho_e$(PDADMAC) $= 9.95 \; 10^{10}$ cm$^{-2}$, $\rho_e$(PAA) $= 9.89 \; 10^{10}$ cm$^{-2}$, and $\rho_e$(H$_2$O) $= 9.46 \; 10^{10}$ cm$^{-2}$. Since the scattering of a given object within the solution is proportional to ($\rho_e$(object)-$\rho_e$(solvent))$^2$, the scattering of the coated nanoparticles in water is completely dominated by the γ-Fe$_2$O$_3$ core due to contrast reasons ( ($\rho_e(\gamma$-Fe$_2$O$_3)$-$\rho_e$(H$_2$O))$^2$/($\rho_e$(PDADMAC)-$\rho_e$(H$_2$O))$^2$ ~ 4000 and ($\rho e(\gamma$-Fe$_2$O$_3)$-$\rho e$(H$_2$O))2/($\rho_e$(PAA)-$\rho_e$(H$_2$O))$^2$ ~ 5000).

**Dynamic light scattering**
Dynamic light scattering (DLS) experiments were carried out using a Zetasizer Nano ZS instrument from Malvern. The instrument uses monochromatic light with a 633 nm wavelength provided by a He-Ne laser at a fixed angle of 173°. The concentration of the nanoparticles (0.0025 v/v% $\gamma$-Fe$_2$O$_3$/Z-PDADMAC) was dilute to avoid any concentration effects (inter-particle interactions) and determine the true self-diffusion coefficient/hydrodynamic size. $Z_{av}$ is the Intensity weighted harmonic mean size using cumulant analysis. In Figure 1 State diagram of the PAA@$\gamma$-Fe$_2$O$_3$ nanoparticles/PDADMAC system, $Z_{av}$ is an average of three measurements of two minutes each. When the hydrodynamic sizes were above a micron (in aggregated systems) and in turn affected by gravity (sedimentation effects), no attempts are made to accurately predict the size distributions.

**Optical microscopy**:
Optical microscopy was used with respective magnification lenses of 10X, 20X, 60X. The samples were deposited on a transparent glass slide, which was furthermore sealed to reduce the evaporation.

# Results and discussion

**Probing the desalting/resalting transition by light scattering: highlighting the formation of persistent aggregates**

We firstly established the state diagram of the system and determined the critical ionic strength $I_c$ by combining visual observation and DLS. The dormant stock solution of 0.025 v/v% $\gamma$-Fe$_2$O$_3$ +

Z-PDADMAC + 1 M NH$_4$Cl was diluted by a factor 10 to lower the salinity down to the targeted ionic strength I$_{target}$. Two dilution pathways were used, respectively fast (or quenching) and drop-by-drop mixing where water is progressively added to the dormant solution (see Material and methods). For both desalting pathways, different charge ratios (Z = 0.04, 0.2, 1) were explored over targeted ionic strength I$_{target}$ ranging from I = 0.1 to 1 M NH$_4$Cl.

It immediately appears that: *(i)* there exist two clear-cut regimes depending on I$_{target}$ with a critical threshold I$_c$ located at ~ 0.62 M; *(ii)* the state diagram does not depend on the pathway, since samples appear similar for a given set of conditions (Z, I) whether desalting is achieved by quenching or drop-by-drop mixing. For I$_{target}$ ≥ I$_c$, samples appear clear and limpid with an orange colour due to the presence of $\gamma$-Fe$_2$O$_3$ maghemite nanoparticles and their aspect does not evolve with time. Complexation does not occur, as shown by DLS (Figure 1.a). Indeed, the hydrodynamic size does not evolve with a value of 32 nm similar to the individual PAA@$\gamma$-Fe$_2$O$_3$ nanoparticles.

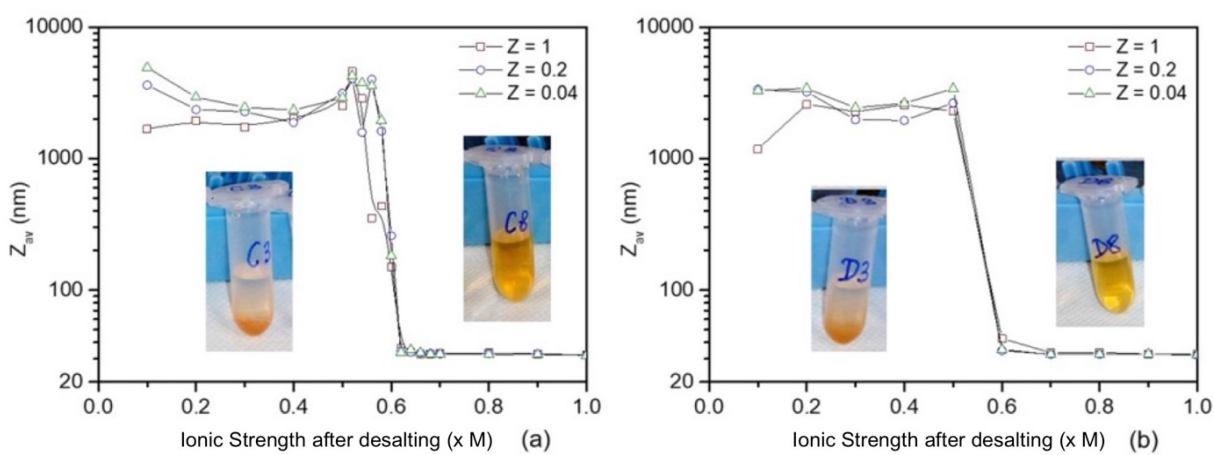

**Figure 1.** State diagram of the PAA@$\gamma$-Fe$_2$O$_3$ nanoparticles/PDADMAC system upon desalting when starting from an ionic strength of 1M NH$_4$Cl for 2 desalting pathways and various Z: (a) Quenching pathway. Evolution of hydrodynamic radius of size complexes, as measured by DLS, versus ionic strength I$_{target}$; Pictures illustrate the macroscopic aspect of samples below I$_c$ (left) and upon the threshold I$_c$ (right). (b) Drop by drop pathway. Evolution of hydrodynamic radius of size complexes, as measured by DLS, versus ionic strength I$_{target}$; Pictures illustrate the macroscopic aspect of samples below I$_c$ (left) and upon the threshold I$_c$ (right).

Conversely, as soon as the critical ionic strength I$_c$ is crossed, complexes are formed whose size grows exponentially with time, as shown by the kinetic evolution of the hydrodynamic radius determined by time resolved DLS measurements (see figure 2.a). Such evolution does not depend on I$_{target}$. The initial stage complexation kinetics was quite fast with a hydrodynamic radius already 10 times larger than individual PAA@$\gamma$-Fe$_2$O$_3$ nanoparticles after just one minute, whatever I$_{target}$. Once the complexes reached a sufficiently large size prone to gravity forces, they started to sediment. Visually, macroscopic samples looked homogeneous up to 30 minutes and progressively started sedimenting afterwards. The characteristic sedimentation time may slightly vary from one sample to another with typical values ranging from 1h to 5h. After 24h, the sedimentation process is over with samples presenting a two-phase system with a clear supernatant and a solid-like precipitate at the bottom of the vial. We characterized then such steady state to establish the state diagram of the system presented in Figure 1. The samples were slightly shaken by hand to get a homogeneous solution just before starting DLS measurements, which enables proper measurements since the measurement time is shorter than the characteristic sedimentation time. The transition threshold is very sharp with onset located at $I_{target}$ = 0.62 M, an ionic strength for which the electrostatic interaction is highly screened since the corresponding Debye length is ~ 0.4 nm. For 0.58 M ≥ $I_{target}$ > 0.62 M, the hydrodynamic radius of the complexes sharply increased when decreasing $I_{target}$ and reached typical values larger than a few hundreds of nm for $I_{target}$ < 0.58 M. The value of $I_c$ is neither influenced by Z (see Figure 1.b) nor by the formulation pathway as both fast mixing and drop by drop mixing give identical results. Such $I_c$ is much larger than those measured in the case of the pure PAA-PDADMAC polyelectrolyte system with the same respective chains $M_w$ and located at 0.3M.[23] The difference arises from the fact that free PAA chains have to pay greater entropic penalty to form a complex with PDADMAC, whereas PAA chains coated on $\gamma$-$Fe_2O_3$ nanoparticles have already lost most of their conformational entropy when adsorbed onto the nanoparticles and can thus easily form complexes with PDADMAC chains upon desalting. Such $I_c$ is however very close to those obtained on the same system when the desalting transition was crossed by dialysis ($I_c$ = 0.54 M).[36]

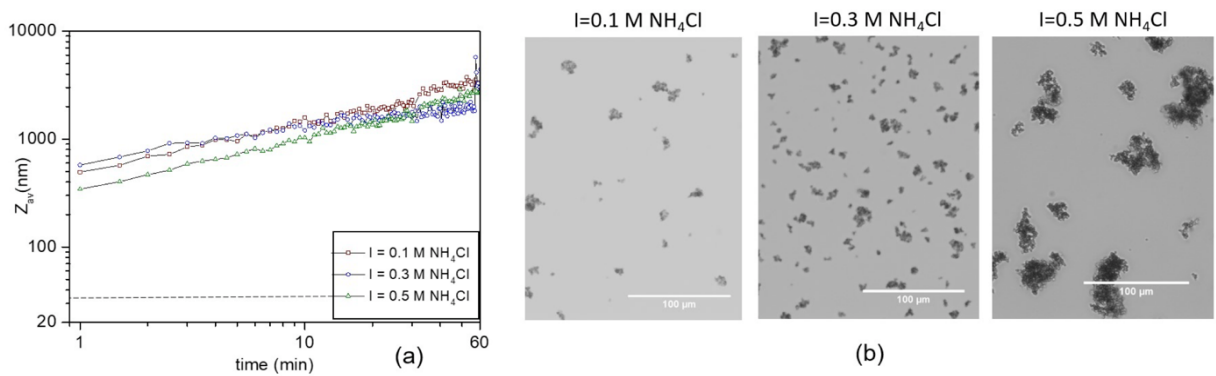

**Figure 2.** Evolution of the aggregates morphology after desalting in the case of the fast mixing pathway at Z = 0.2 for various $I_{target}$: (a) kinetic evolution of the size by DLS; the dotted line recalls the hydrodynamic radius of single PAA@$\gamma$-$Fe_2O_3$ nanoparticles. (b) Optical images of the aggregates are obtained after quenching time of 72h.

In order to assess if the transition is driven by either a liquid-solid (precipitation) or a liquid-liquid transition, *i.e.* a complex coacervation, we imaged the aggregates by optical microscopy. Images were taken 3 days after mixing (figure 2.b). They appeared as dense polymorph precipitates and not at all to spherical objects with an overall size that slightly increased from a few tens of  m at $I_{target}$ = 0.1M up to a few hundreds of  m at $I_{target}$ = 0.5M. This suggests at first sight that the system undergoes precipitation rather than complex coacervation.

Interesting results were obtained when crossing back the transition by resalting the system to the nominal 1 M ionic strength (Figure 3). The behaviour of the system was indeed dependent on the "quenching time" defined as the time elapsed between the desalting and resalting steps, or in other words the resting time at $I_{target}$. For short quenching times (from a few minutes up to ~1 hour), the complexes re-dissociate upon salt addition as shown by time-resolved DLS experiments (Figure 3.a), which suggests that the desalting transition is reversible. However, when the quenching time was much longer (~days), persistent aggregates were still visible (Figure 3.b). There is thus in the system an ageing time during which aggregates tend to reorganise toward more stable states after quenching. Moreover, $I_{target}$ strongly influenced the stability of such permanent aggregates. Indeed, Figure 3.b presents the kinetic evolution of the size after resalting for different $I_{target}$: the closer $I_{target}$ to $I_c$ is, the more difficult it is to dissolve the aggregates. Moreover, for a given $I_{target}$, the overall size of the aggregates increases with the quenching time (figure 3.c). This result is at first sight rather counterintuitive, as one would expect that the reversibility of the transition would be easier when electrostatic interactions are highly screened close to the threshold. Actually, it comes from the way $I_{target}$ tunes the internal morphologies of the complexes, as will be shown further down with the help of SAXS experiments.

Interestingly, at very long quenching times (3 months), the dissolution rate for $I_{target}$ = 0.5 was slow enough to enable a visualization of the complexes immediately after resalting by optical microscopy (figure 3.d). Contrary to what was observed just after quenching, some aggregates appeared spherical, which suggests that they form a dense liquid phase. It is then likely that the system undergoes complex coacervation rather than precipitation upon quenching in the vicinity of $I_c$ well in line with a weaker electrostatic interaction strength. Coacervates are usually very viscous[48,49] and it takes them a very long time to reach a spherical droplet shape, a process that is made possible by the reorganizations that occur upon ageing. The behaviour of the system recalls thus those already observed in oppositely charged polyelectrolyte complexes where precipitation occurs far from $I_c$ and coacervation close to $I_c$[50].

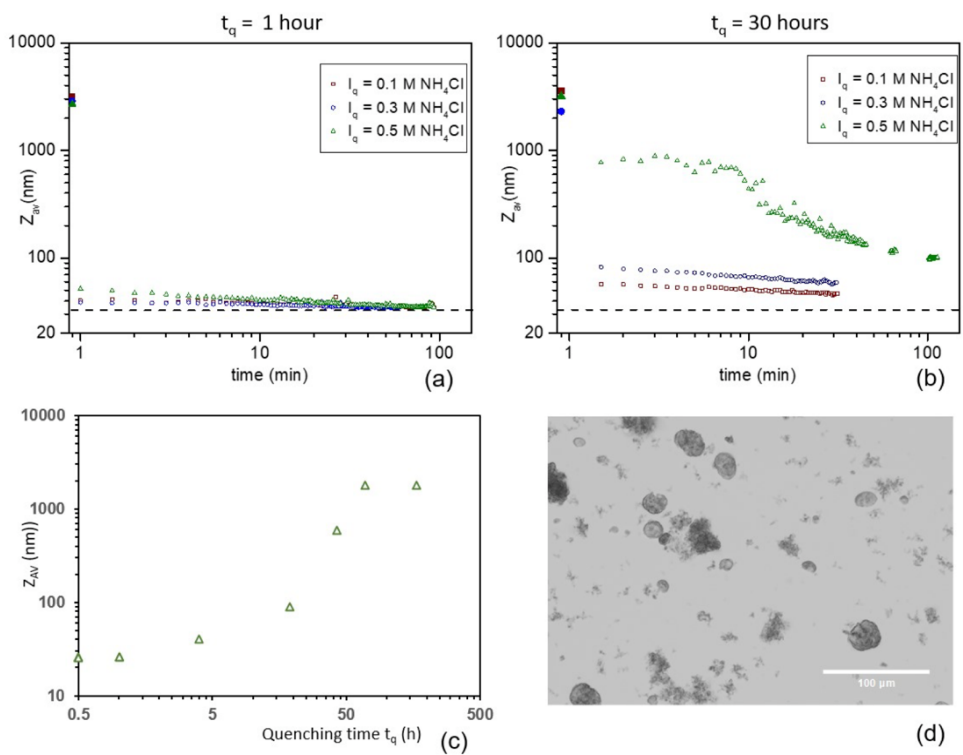

**Figure 3.** Evolution of the aggregate morphology after resalting along the desalting/resalting pathway in the case of the fast mixing pathway at Z = 0.2 for various $I_{target}$. (a) and (b): kinetic evolution of their average size by DLS, respectively when the salting step has been achieved 1h after initial desalting (a) or 30 h after initial desalting (b); the dotted line recalls the hydrodynamic radius of single PAA@$\gamma$-Fe$_2$O$_3$ nanoparticles. (c) Evolution of the average size as function of quenching time for I$_{target}$ = 0.5M. (d) Optical images of the aggregates obtained immediately after resalting after an elapsed quenching time of 3 months for I$_{target}$ = 0.5M.

**Probing the structure of aggregates by SAXS: I$_{target}$ tunes their inner structure and in turn their ability to re-dissociate**

DLS time-resolved results suggested that depending on I$_{target}$, the inner morphology of the complexes presented some differences. This prompted us to determine their inner structure at the SAXS scale. Since both Z and the formulation pathway have only a very weak influence on the complex formation (Figure 1), we decided to focus only on the quenching pathway at Z=0.2. In the SAXS experiments, only the $\gamma$-Fe$_2$O$_3$ core nanoparticle contributes to the scattering due to contrast reasons (see materials and methods). The scattering intensity writes then:

$$I(q)(cm^{-1}) = \Phi \Delta \rho^2 P(q) S(q) \quad (1)$$

where $\Phi$ is the volume fraction of $\gamma$-Fe$_2$O$_3$ nanoparticles, $\Delta \boldsymbol{\rho}$ the difference of electronic scattering length density between $\gamma$-Fe$_2$O$_3$ nanoparticles and water, *P(q) and S(q)* the form and the structure factor of $\gamma$-Fe$_2$O$_3$ nanoparticles respectively.

SAXS experiments were performed 24h after quenching. When the dispersion have phase-separated below I$_c$, the dense phase present at the bottom of the vial was studied. Figure 4.a shows a series of SAXS curves measured for various I$_{target}$ ranging from 0.1 M to 0.65M together with the form factor of individual $\boldsymbol{\gamma}$-Fe$_2$O$_3$ nanoparticles. At 0.65M, the SAXS spectra superimpose with the nanoparticles form factor because complexation has not yet occurred. For all others I$_{target}$, the scattering is strongly modified and present two main features : *(i)* at low q, it decays like q$^{-4}$ (Porod law) in line with the formation of large and dense complexes (seen by DLS); *(ii)* at intermediate q, there is a strong correlation peak at ~ 0.6 nm$^{-1}$ that corresponds to the mean distance between nanoparticles within the aggregates. The intensity and q-position of such peak q$^*$ strongly depend on I$_{target}$, as highlighted in Figure 4.b that presents the structure factor of the nanoparticles within aggregates.

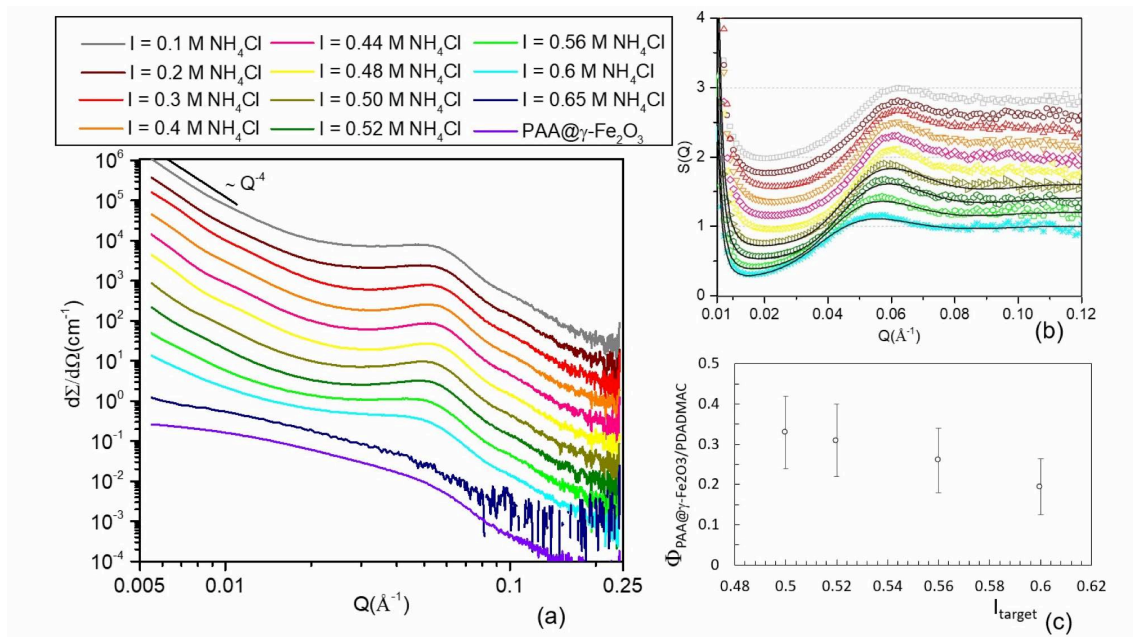

**Figure 4.** SAXS scattering curves of the PAA@$\gamma$-Fe$_2$O$_3$ nanoparticles/PDADMAC system upon desalting from 1M NH$_4$Cl by quenching at Z = 0.2. (a) SAXS data in log-log scale at different ionic strengths compared with the form factor of pure PAA@$\gamma$-Fe$_2$O$_3$ nanoparticles (purple). The scattering of pure PAA@$\gamma$-Fe$_2$O$_3$ nanoparticles is in absolute scale. As the nanoparticle concentration may change from one sample to another in presence of salt due to sedimentation, all curves have been scaled on the form factor at large q. They were thene shifted from one to another by a factor 3.33 for clarity; (b) Structure factor of complexes for $I_{target} \leq 0.6$ M , as obtained by dividing the scattering intensity by the form factor of individual PAA@$\gamma$-Fe$_2$O$_3$ nanoparticles. All curves are separated one from each other by 0.2 for clarity. The continuous black lines are fits by a Percus-Yevick structure factor. The error bars are smaller than the symbols. (c) Evolution of the compactness of aggregates as function of $I_{target}$ in the vicinity of $I_c$.

Two regimes can be identified. When $I_{target}$ is in the vicinity of $I_c$ ($I_{target} > 0.48$), the inner structure of the aggregates strongly changes upon a slight variation of $I_{target}$. The correlation peak is initially very broad at 0.6 M just below $I_c$. As one moves away from the transition by decreasing $I_{target}$, $q^*$ is shifted towards larger q (from 0.54 nm$^{-1}$ at 0.6M up to 0.615 nm$^{-1}$ at 0.48 M, which corresponds to a decrease of ~ 15% of the mean distance between nanoparticles) while its height $S(q^*)$ increases (see Figure S2 in supplementary information). It suggests that the complexes contain a certain amount of water at 0.6 M and become progressively denser and more compact and then less solvated when moving away for the transition. In order to account quantitatively for such change of compactness in this regime of $I_{target}$, we have used a Percus-Yevick structure factor[51] in order to obtain the total inner volume fraction of objects of nanoparticles and PDADMAC within the aggregates. Such structure factor describes hard-spheres systems and usually reproduces well the scattering features of the inner structure of dense aggregates of spheres with only two parameters: the hard sphere radius $R_{HS}$ and volume fraction. Figure 4.c shows the results obtained from the fits of the PY structure factors for a constant $R_{HS}$ of 5.6±0.1nm. The parameters of the fits are given in supplementary information. $R_{HS}$ is slightly larger than the individual $\gamma$-Fe$_2$O$_3$ nanoparticles (4.7nm) because it takes into account the overall size of the nanoparticles made of a $\gamma$-Fe$_2$O$_3$ core, a functional adsorbed PAA layer and the PDADMAC chains. $\Phi_{PAA@\gamma\text{-Fe2O3/PDADMAC}}$ decreases continuously from 0.33 at 0.5 M down to 0.19 at 0.62 M.

When $I_{target}$ is far from $I_c$ ($I_{target} \leq 0.48$), $q^*$ stays constant at 0.615 nm$^{-1}$ (10.2 nm in direct space) showing that compactness has reached a maximal value. The correlation peak broadens and its height $S(q^*)$ slightly decreases when $I_{target}$ decreases, showing that the system is less organized (see Figure 4.b.). This is probably due to a system that freezes during complexation with a magnitude increasing inversely with $I_{target}$ in line with a stronger electrostatic interaction. It turns out that nanoparticles have more freedom to move and reorganise within the complexes in the first regime ($I_{target} > 0.48$) since their compactness is still far from its maximal value. It does not however hamper the integrity of such aggregates, which remain persistent with time. The final aggregates morphology is thus tuneable with a very slight modification of the ionic strength $I_{target}$.

DLS data (Figure 2.a) have shown that these complexes are formed within short time scales (less than 1h, mostly few minutes or less). In order to assess if their inner structure does evolve on a much longer time scale, we have performed SAXS experiments from 24h to 30 days after quenching for different ionic strengths. We have found that the complex inner structure does not evolve with time as SAXS curves were similar up to 1 month after quenching (see figure S3 in Supplementary information that compares data for several $I_{target}$ ranging from 0.1 M to 0.56 M).

The kinetics of dissolution of complexes after resalting(crossing back the threshold to 1M NH$_4$Cl) was probed for 3representative $I_{target}$ (0.1 M, 0.3 M and 0.5 M, Figure 5.a to Figure 5.c). The quenching time was set to 10 days, a much larger time that the characteristic time used for DLS observations. The measurements were performed just after the resalting step. For $I_{target}$ = 0.1 M, dissolution is almost immediate as the SAXS curve superimposes well with the form factor after 45 minutes only with no observed evolution afterwards. Only a slight aggregation is present in the system, as shown by a very small increase of the scattering intensity at the lowest q. For $I_{target}$ = 0.3 M, where the system is slightly less frozen, the behaviour is very close to those at 0.1 M, even though some aggregates still persist and do not disassociate with time, in accordance with DLS results. On the contrary, for $I_{target}$ = 0.5 M, when the complexes contain more water, the morphology of the complexes evolves progressively over hours with aggregates still present after 1 day. The dissolution of the structures with a quenching time of 10 days is thus very slow. The way the correlation peak evolves upon dissolution at $I_{target}$ = 0.5 M is also informative: it broadens with dissolution time but its q-position does not change. This shows that the internal morphology of the aggregates remains the same after resalting, the broadening arising from the decrease of the overall size of the aggregates with time. This means that the dissolution mechanisms are probably driven by a progressive release of the nanoparticles from the outer surface of aggregates.

Such kinetic SAXS experiments demonstrate that the inner structure of the complexes drives their ability to reorganize toward more stable states. Indeed, for strong electrostatic interactions, frozen aggregates are formed. The freezing process has thus trapped the system in a long-lived configuration as far from its minimal free energy as $I_{target}$ is low. The system cannot then reorganize, evolve or age after quenching while it can recover easily its initial state upon desalting, which makes the transition almost reversible. At 0.1 M, dissolution is fully reversible while a very limited amount of aggregates persists at 0.3M. It is likely that the system has reached a more stable configuration at 0.3M than at 0.1M. On the contrary, for weak electrostatic interactions, when $I_{target}$ is close to $I_c$, the compactness of the aggregates is not maximal. Reorganizations can then occur with time and may follow entropy driven rearrangements toward kinetically trapped metastable assemblies, which have a lower free energy than the states formed just after quenching. Aggregates reach then a configuration where they are more stable and therefore very difficult to dissolve, which increases consecutively their lifetime after desalting. Permanent metastable aggregates can thus be obtained upon resalting if the time elapsed after the initial desalting is long enough. Figure 6 presents a schematic overview of the evolution of the system.

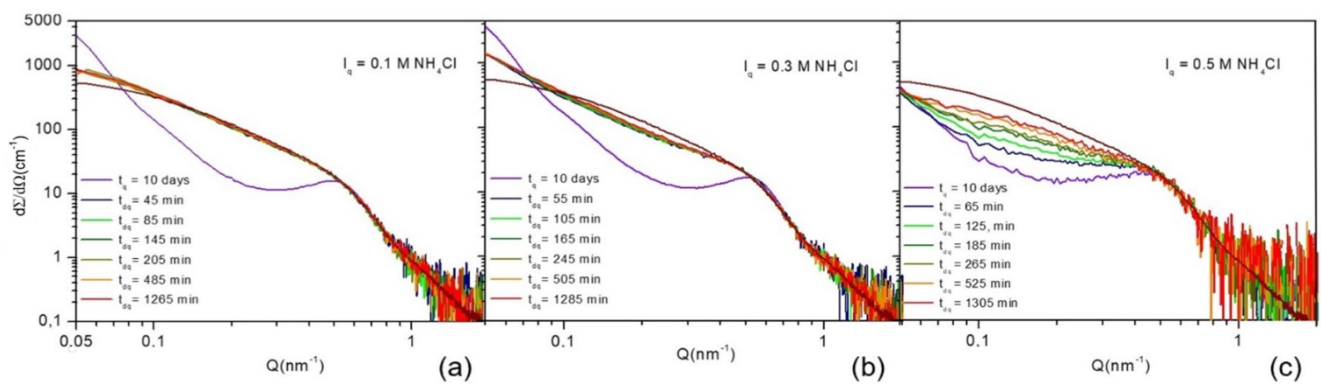

**Figure 5**. Kinetic evolution of the inner morphology of the complexes by SAXS after salting back (dequenching) 10 days after quenching along the desalting/salting pathway in the case of the quenching mixing pathway at Z = 0.2 for various $I_{target}$. (a) $I_{target}$ = 0.1 M; (b) $I_{target}$ = 0.3 M; (c) $It_{arget}$ = 0.5 M. SAXS curves are compared to the form factor of individual PAA@$\gamma$-Fe$_2$O$_3$ nanoparticles (brown curve).

## Conclusion

In summary we have shown that, the desalting/resalting pathway enable to form aggregates with different dissolution rates: they dissolve fast upon resalting at low ionic strength ($I_{target}$ = 0.1 M) far below $I_c$, whereas they are more stable and dissolve very slowly at higher ionic strength ($I_{target}$ = 0.5 M) close to $I_c$. While such results may appear counterintuitive at first sight (stronger electrostatic interactions lead to less stable aggregates), they can be rationalized by the way the freezing of the system upon quenching restrict any further entropy driven rearrangements within complexes. The re-dissolution of complexes upon resalting is indeed driven by the internal morphology of complexes which is set by $I_{target}$. Internal rearrangements within the aggregates are the largest in the vicinity of $I_c$ when the volume fraction of nanoparticles within the aggregates is the lowest (Figure 4.c). Within this limited range of $I_{target}$ (0.5M to 0.6 M NH$_4$Cl), the reorganization rate within the aggregates is sufficiently large to enable the formation of very stable aggregates resistant to dissolution. The desalting/resalting pathway is then ideal to form metastable states with various morphologies, sizes, dissolution rates and lifetimes. In particular, it can be used advantageously (i) when stabilization of specific nanoparticle co-assemblies is required in conditions where they are not spontaneously associated at equilibrium *at* very high ionic strength; (ii) or to slowly deliver drugs or precursors in chemical reactions.

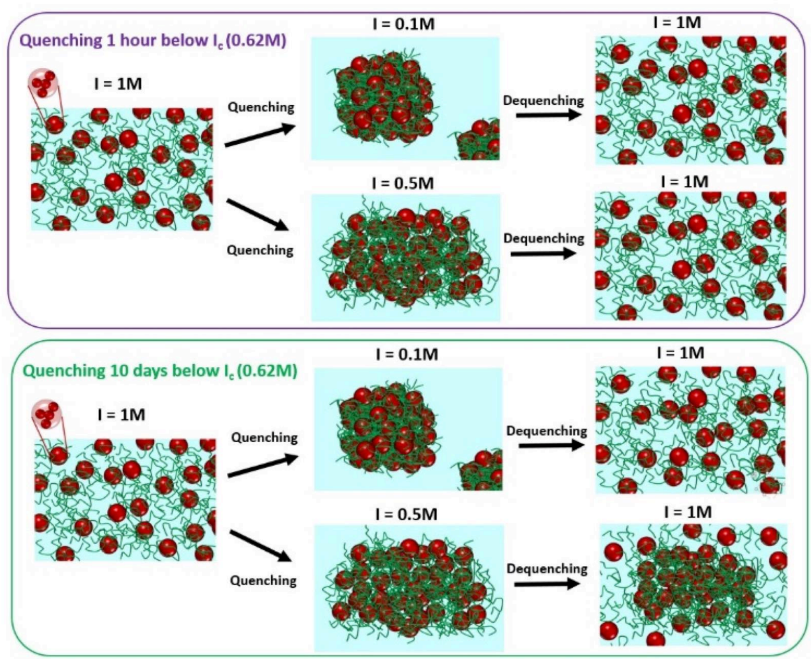

**Figure 6.** Overview of the evolution of the system along the desalting/salting pathway for different $I_{target}$ and different elapsed time between quenching and dequenching. $I_c$ points out the critical ionic strength at which electrostatic complexation occurs ($I_c$ =0.62 M). The overall size of aggregates are not at scale.

## Conflicts of interest
There are no conflicts to declare.

## Acknowledgements

This research was supported by Agence Nationale de la Recherche (PANORAMA project, ANR-13-BS08-0015).